\documentclass[twocolumn]{aastex701}

\usepackage{amsmath,amssymb,amsfonts}
\usepackage{booktabs}
\usepackage{microtype}
\usepackage{graphicx}
\usepackage{url}

\newcommand{\target}{HD~70510}
\newcommand{\tic}{TIC~388351327}
\newcommand{\dnu}{\Delta\nu}

\newcommand{\teff}{T_{\rm eff}}

\shorttitle{Asteroseismology of the Gaia SB2 HD 70510}
\shortauthors{Lv et al.}

\begin{document}

\title{Asteroseismology of the Gaia SB2 HD 70510: From Radial Modes to Component Masses}

\author[orcid=0000-0001-6354-1646]{Chenglong Lv}
\affiliation{Xinjiang Astronomical Observatory, Chinese Academy of Sciences, Urumqi, Xinjiang 830011, People's Republic of China}
\affiliation{Instituto de Astrof\'isica de Andaluc\'ia - CSIC, 18008 Granada, Spain}
\email[show]{lvchenglong@xao.ac.cn}

\author[orcid=0000-0003-1845-4900]{Ali Esamdin}
\affiliation{Xinjiang Astronomical Observatory, Chinese Academy of Sciences, Urumqi, Xinjiang 830011, People's Republic of China}
\affiliation{School of Astronomy and Space Science, University of Chinese Academy of Sciences, Beijing 100049, People's Republic of China}
\email{aliyi@xao.ac.cn}

\author[orcid=0000-0001-5066-5682]{Shuguo Ma}
\affiliation{Xinjiang Astronomical Observatory, Chinese Academy of Sciences, Urumqi, Xinjiang 830011, People's Republic of China}
\email{mashuguo@xao.ac.cn}

\author[orcid=0000-0001-8677-2391]{Shahidin Yaqup}
\affiliation{Xinjiang Astronomical Observatory, Chinese Academy of Sciences, Urumqi, Xinjiang 830011, People's Republic of China}
\affiliation{School of Astronomy and Space Science, University of Chinese Academy of Sciences, Beijing 100049, People's Republic of China}
\email{xiayiding@xao.ac.cn}

\author[orcid=0000-0002-2463-4943]{Haozhi Wang}
\affiliation{Chinese Academy of Sciences South America Center for Astronomy, Chile}
\email{wanghaozhi@xao.ac.cn}

\author[orcid=0000-0001-9600-963X]{Ayzada Jmahali}
\affiliation{Xinjiang Astronomical Observatory, Chinese Academy of Sciences, Urumqi, Xinjiang 830011, People's Republic of China}
\affiliation{School of Astronomy and Space Science, University of Chinese Academy of Sciences, Beijing 100049, People's Republic of China}
\email{Ayzada@xao.ac.cn}

\author[orcid=0009-0000-2772-2677]{Jianwei Yao}
\affiliation{Xinjiang Astronomical Observatory, Chinese Academy of Sciences, Urumqi, Xinjiang 830011, People's Republic of China}
\affiliation{School of Astronomy and Space Science, University of Chinese Academy of Sciences, Beijing 100049, People's Republic of China}
\email{yaojianwei@xao.ac.cn}

\author[orcid=0009-0008-7094-1304]{Kedieryemu Kamili}
\affiliation{School of Physical Science and Technology, Xinjiang University, Urumqi 830017, People's Republic of China}
\email{kedieryemu@xao.ac.cn}

\begin{abstract}
Double-lined spectroscopic binary (SB2) orbits give mass ratios and projected component masses, but the absolute masses depend on the unknown inclination. HD~70510 is a Gaia DR3 SB2 in which regular $\delta$~Scuti pulsations constrain the properties of one component. We analyze nine sectors of TESS photometry and combine stellar evolution and pulsation modelling with a two-star spectral energy distribution (SED) fit to the unresolved Gaia and 2MASS photometry. Five high-signal-to-noise peaks form a sequence consistent with the radial fundamental mode and four consecutive overtones. Two stable pulsation clocks show the sign and amplitude expected from the orbital motion of the Gaia primary, identifying it as the pulsator. Combining the radial-mode and binary-SED fits with the Gaia SB2 orbit gives $M_1=1.594^{+0.026}_{-0.026}\,M_\odot$, $R_1=1.482^{+0.014}_{-0.012}\,R_\odot$, and an age of $171^{+112}_{-99}$~Myr. We obtain an orbital inclination of $i=82.1^{+3.5}_{-3.4}$~deg and a companion mass of $M_2=1.121^{+0.029}_{-0.028}\,M_\odot$. These intervals are statistical and remain conditional on the adopted radial-mode identification and stellar and binary-SED models. The primary lies close to the zero-age main sequence, while the roughly 6000-K companion contributes 14--28\% of the unresolved optical-to-near-infrared flux. HD~70510 shows how regular pulsations can provide component masses for non-eclipsing TESS binaries with Gaia DR3 SB2 orbital solutions.
\end{abstract}


\section{Introduction}\label{sec:intro}

The pulsation spectra of $\delta$~Scuti stars contain low-order pressure modes whose frequencies are affected by mode selection, rotation, and stellar structure. Secure mode identification is therefore a major challenge in quantitative modelling. The Transiting Exoplanet Survey Satellite (TESS) revealed that some young $\delta$~Scuti stars show regular high-frequency ridges that can be organized into radial and nonradial sequences \citep{Bedding2020}. The ridge spacing, $\dnu$, traces mean density, although these low-order modes do not follow the solar-like asymptotic relation exactly \citep{GarciaHernandez2017,Murphy2023,Lv2024}. Consecutive radial modes and their period ratios therefore provide useful constraints for individual $\delta$~Scuti stars and samples of regular pulsators \citep{Lv2022,Lv2023,Lv2026a}.

Binary information can reduce degeneracies left by a single-star analysis. The two velocity amplitudes of a double-lined spectroscopic binary (SB2) determine its mass ratio and projected component masses, but not the inclination or absolute masses. An independent mass for either component fixes the mass scale. Stable pulsations also connect the photometric and spectroscopic descriptions because orbital light-travel time changes their observed phases \citep{Murphy2014,Murphy2018,Lv2026c,Lv2026b}. In KIC~9773821, for example, $\delta$~Scuti pulsation timing was combined with red-giant asteroseismology and spectroscopy \citep{Murphy2021}. The Gaia Data Release 3 (DR3) non-single-star (NSS) catalogue makes similar tests possible for many more binaries \citep{GaiaMultiplicity,Beck2024}.

\target\ (\tic) has both types of information. \citet{Bedding2020} identified its regular high-frequency pattern, and \citet{Murphy2023} measured $\dnu=7.24$~d$^{-1}$, proposed a radial fundamental mode near 21.68~d$^{-1}$, and reported $v\sin i=94\pm10$~km~s$^{-1}$, although the latter may be affected by the binary nature of the source. In a study of 60 regular pulsators, \citet{Panda2024} obtained a single-star solution near $1.65\,M_\odot$, $Z=0.018$, and 90~Myr. Independently, Gaia DR3 classified the source as an SB2 with a 42.23-d eccentric orbit \citep{GaiaMultiplicity}. Those studies established the pulsation pattern and binary orbit, but did not identify the pulsating spectroscopic component or remove the companion's contribution from the unresolved photometry used to constrain the pulsator.

We first use pulsation timing to identify which spectroscopic component pulsates and to search for a second clock from the companion. We then model the radial-mode sequence and a physical two-star spectral energy distribution (SED), and combine the resulting stellar constraints with the Gaia orbit. The same orbital solution supplies the mass ratio used in the SED and the projected masses used to infer the inclination. This yields a component-specific characterization of both stars in a non-eclipsing system and provides a framework that can be applied to other regular TESS pulsators with Gaia NSS SB2 solutions.

\section{Observations and Frequency Spectrum}\label{sec:data}

\subsection{TESS photometry}

We used the publicly available two-minute light curves produced by the TESS Science Processing Operations Center (SPOC) for Sectors 8, 9, 34--36, 61, 63, 88, and 90 \citep{Ricker2015}. After rejecting flagged and non-finite cadences and normalizing each sector, 147,976 measurements remain over an elapsed baseline of 2257~d. All amplitudes are expressed in mmag. No eclipses are evident in the available TESS photometry.

The multi-year gaps provide a long time span for pulsation timing but produce a complicated spectral window if all nine sectors are combined for frequency extraction. We therefore calculated the amplitude spectrum of the consecutive Sectors 34--36 and extracted frequencies with FELIX using iterative prewhitening \citep{Charpinet2010}. At each step, the strongest remaining peak was fitted and removed. The accepted frequencies, amplitudes, and phases were then refined simultaneously. The reported signal-to-noise ratios (S/N) are measured from the local residual spectrum, and the formal uncertainties come from the final multi-sinusoid solution. The unresolved companion and any constant field contamination dilute the observed amplitudes but do not shift the frequency or phase of a coherent pulsation; all amplitudes are therefore quoted relative to the observed system light. The five radial-mode peaks also appear in the three other TESS observing epochs (Sectors 8--9, 61/63, and 88/90). At each epoch, the recovered frequencies agree with the Sectors 34--36 values within one Rayleigh resolution element, $1/T$. The largest offset is 0.0016~d$^{-1}$, equivalent to 0.12 Rayleigh resolution elements. We retain the Sectors 34--36 frequencies because this continuous three-sector light curve has the cleanest spectral window.

The formal frequency uncertainties measure the precision of the light-curve fit, not the accuracy of the stellar models. We therefore adopt model-comparison uncertainties of 0.50, 0.35, and 0.30~d$^{-1}$ for $n=1$, $n=2$, and $n=3$--5, respectively. These scales allow for residual differences between the star and the one-dimensional models. Interpolation and rotational contributions are estimated separately and added in quadrature.

\begin{figure*}
 \centering
 \includegraphics[width=0.97\textwidth]{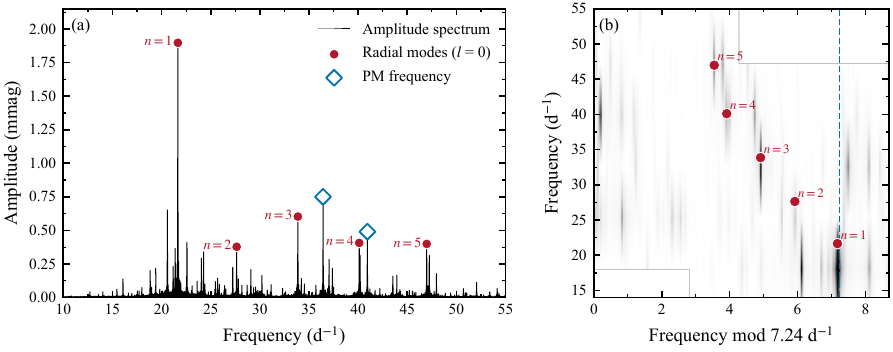}
 \caption{Frequency pattern of \target\ from TESS Sectors 34--36.  (a) Fourier amplitude spectrum in system-light mmag. Red circles mark the five candidate radial modes, and blue open diamonds mark the two isolated frequencies used as pulsation clocks. (b) Échelle diagram constructed with $\Delta\nu=7.24$~d$^{-1}$. The final 20\% of the horizontal axis repeats the start of the next order to show the ridge continuously across the modulo boundary.}
 \label{fig:pulsation}
\end{figure*}

\subsection{The regular p-mode ridge}\label{sec:modes}

Figure~\ref{fig:pulsation} shows five high-S/N peaks at 21.6574, 27.6418, 33.8718, 40.1142, and 46.9868~d$^{-1}$ on the curved low-order ridge obtained by folding the spectrum at $\dnu=7.24$~d$^{-1}$ \citep{Murphy2023}. The four adjacent radial-mode spacings are 5.9844, 6.2300, 6.2425, and 6.8726~d$^{-1}$, showing that the low-order radial sequence is not equally spaced. The consecutive observed period ratios, from $P_2/P_1$ through $P_5/P_4$, are 0.7835, 0.8161, 0.8444, and 0.8537. Together with the location of the lowest-frequency peak, these ratios support its identification as the radial fundamental mode followed by four consecutive overtones. We label them $n=1$--5 and test the complete sequence against the stellar-model grid rather than identifying each peak independently. We test the alternative $n=2$--6 and $n=3$--7 assignments quantitatively in Section~\ref{sec:results}.

\begin{deluxetable*}{cccccc}
\setlength{\tabcolsep}{6pt}
\tablecaption{Adopted Radial-mode Candidates\label{tab:modes}}
\tablehead{\colhead{$n$} & \colhead{$\nu_{\rm obs}$} & \colhead{$\sigma_{\nu,\rm formal}$} & \colhead{$\sigma_{\rm adopt}$} & \colhead{Amplitude} & \colhead{S/N}\\
& \colhead{(d$^{-1}$)} & \colhead{(d$^{-1}$)} & \colhead{(d$^{-1}$)} & \colhead{(mmag)} &}
\startdata
1 & 21.65739 & 0.00001 & 0.50 & $2.222\pm0.004$ & 623\\
2 & 27.64181 & 0.00006 & 0.35 & $0.474\pm0.004$ & 122\\
3 & 33.87178 & 0.00003 & 0.30 & $0.833\pm0.004$ & 214\\
4 & 40.11424 & 0.00006 & 0.30 & $0.479\pm0.004$ & 116\\
5 & 46.98683 & 0.00006 & 0.30 & $0.515\pm0.004$ & 117\\
\enddata
\tablecomments{Frequencies, formal errors, amplitudes, and S/N values come from the Sectors 34--36 solution. The adopted uncertainties are baseline model-comparison floors that allow for unresolved one-dimensional model differences, including the outer-layer treatment. Interpolation and rotation terms are estimated independently and added separately in quadrature in the stellar likelihood.}
\end{deluxetable*}

\section{Pulsation Timing and the Gaia SB2 Orbit}\label{sec:timing}

The Gaia NSS orbit uses 21 epoch radial velocities for each component. It has $P=42.23374\pm0.00535$~d, $e=0.3997\pm0.0071$, $K_1=38.86\pm1.29$~km~s$^{-1}$, and $K_2=54.31\pm0.69$~km~s$^{-1}$. The velocity amplitudes imply $q=M_2/M_1=K_1/K_2\simeq0.716$. They also determine the projected orbital sizes and hence predicted light-time amplitudes of 69.0~s for the primary and 96.4~s for the secondary. The two signals are anti-phased because the stars move on opposite sides of the barycenter. Gaia spectroscopy does not itself determine the orbital inclination. We retain the published correlations among the orbital elements in the mass calculation below; the marginal uncertainties quoted here summarize only the individual catalogue parameters.

Following the phase-modulation method of \citet{Murphy2014,Murphy2018}, we measured pulsation phases in 7-d segments. Tests with 5, 8, and 10~d segments showed that 7~d provides a good balance between timing precision and sampling across the 42-d orbit. The highest-amplitude modes are not necessarily the best clocks: a useful timing frequency must also be isolated and phase stable. The five radial-mode candidates are therefore used in the stellar model but are not necessarily suitable for timing. Among the isolated peaks, 36.43193 and 40.94438~d$^{-1}$ give the smallest delay uncertainties and the most coherent behavior across the TESS campaigns. With the Gaia orbit fixed, their individual light-time amplitudes are $57.1\pm7.0$ and $65.3\pm9.0$~s. Propagating the full Gaia orbital covariance gives a predicted primary amplitude of $69.00\pm2.28$~s. After combining the uncertainties, the two measured amplitudes differ from this prediction by 1.62 and 0.40 standard deviations, respectively. Both are therefore consistent with the Gaia primary.

\begin{figure*}
 \centering
 \includegraphics[width=0.97\textwidth]{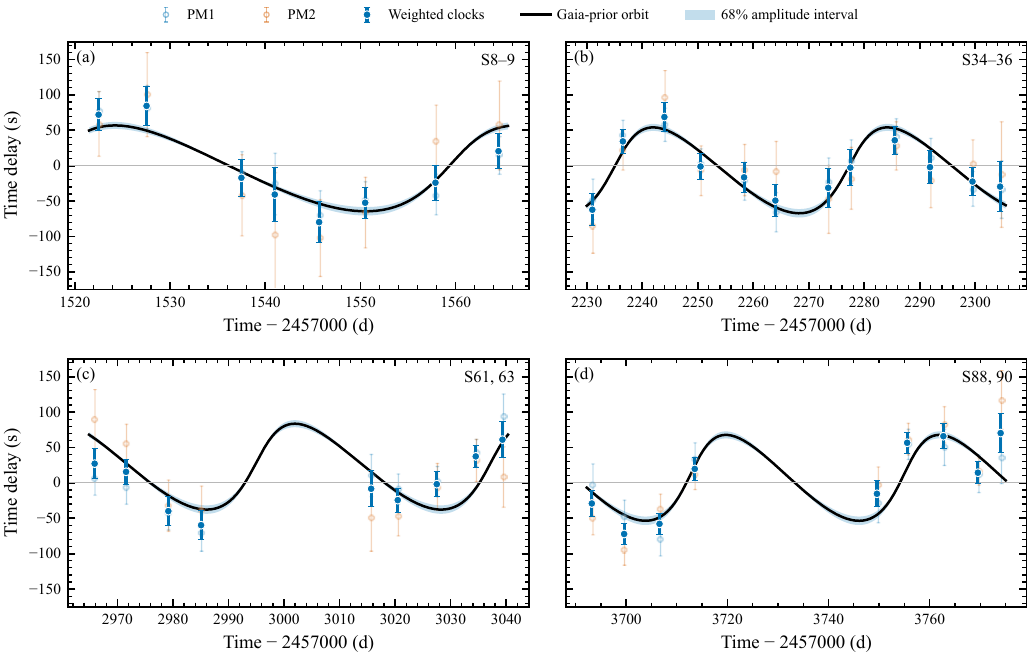}
 \caption{Pulsation delays in the four TESS observing campaigns.  Panels (a)--(d) show Sectors 8--9, 34--36, 61 and 63, and 88 and 90.  Pale blue and orange symbols show the 36.43193 and 40.94438~d$^{-1}$ clocks, and filled blue symbols are their inverse-variance combinations.  The black curve is the Gaia-prior solution; the pale-blue band gives its 68\% light-time-amplitude interval.  Each campaign has an independent zero point because the pulsation cycle count cannot be retained across the multi-year gaps.}
 \label{fig:timing}
\end{figure*}

Figure~\ref{fig:timing} compares three fits to the pulsation delays, and Table~\ref{tab:timing} summarizes their roles. Fixing the Gaia orbital elements gives $a_1\sin i/c=60.6\pm5.7$~s, consistent with the predicted 69.0~s. Allowing the orbit to vary under the Gaia priors gives $65.9\pm5.9$~s. A phase-modulation-only (PM-only) fit also favors a period near 42~d and an amplitude near 66~s, but the separated TESS observing windows leave aliases near 40 and 45~d and only weakly constrain the eccentricity and argument of periastron. Because orbital coherence was considered when selecting the final clocks, the timing analysis is a component-association test rather than a blind orbital discovery. It shows that the observed $\delta$~Scuti delays follow the Gaia primary, but does not improve the Gaia spectroscopic orbit.

\begin{deluxetable*}{lcccl}
\tabletypesize{\small}
\setlength{\tabcolsep}{20pt}
\tablecaption{Pulsation Timing Orbit Fits\label{tab:timing}}
\tablehead{
\colhead{Model} &
\colhead{$a_1\sin i/c$ (s)} &
\colhead{$P_{\rm orb}$ (d)} &
\colhead{$e$} &
\colhead{Role}
}
\startdata
Gaia fixed & $60.6\pm5.7$ & 42.23374 (fixed) & 0.3997 (fixed) & Light-time amplitude \\
Gaia prior & $65.9\pm5.9$ & $42.221\pm0.005$ & $0.400\pm0.007$ & Component association \\
PM-only & $66.1^{+11.0}_{-8.6}$ & $\sim42$ & \nodata & Alias diagnostic
\enddata
\tablecomments{Quoted uncertainties summarize posterior 16th--84th percentiles. The PM-only period posterior has a dominant mode near 42~d and secondary modes near 39.9 and 44.7~d; a single period interval and eccentricity are therefore not reported for this diagnostic fit.}
\end{deluxetable*}

\begin{figure}
 \centering
 \includegraphics[width=0.45\textwidth]{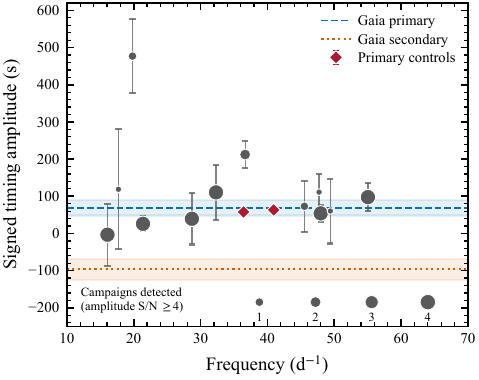}
\caption{Search for a coherent pulsation clock from the companion. Circles give the signed light-time amplitudes and their uncertainties, and symbol area gives the number of TESS campaigns in which each frequency is detected with amplitude S/N $\geq4$. The blue dashed and orange dotted lines mark the Gaia primary and anti-phased secondary predictions, respectively; the shaded regions indicate $\pm30\%$ around those amplitudes. Red diamonds show the two recovered primary controls. None of the 12 secondary-search frequencies meets all companion-clock criteria.}
 \label{fig:secondary}
\end{figure}

The Gaia-fixed fit measures only the light-time amplitude, the Gaia-prior fit tests consistency with the spectroscopic orbit, and the PM-only fit shows the orbital constraints and period aliases present in the TESS timing data alone. We use the Gaia-prior solution for visualization, while the dynamical analysis uses the Gaia NSS orbit and its full covariance.

We next searched for a stable clock following the predicted 96.4-s motion of the secondary. Twelve isolated frequencies between 10 and 70~d$^{-1}$ were tested using 7-d phase measurements across the nine TESS sectors. A candidate was required to be detected with amplitude S/N $\geq4$ in at least three campaigns, have the opposite timing sign to the primary, and have a timing amplitude consistent with the Gaia prediction. As a validation test, the two known primary clocks were excluded from the secondary-candidate list and analyzed in the same way. Both were recovered with the correct sign and amplitudes of $57.8\pm7.4$ and $63.2\pm9.7$~s. Figure~\ref{fig:secondary} shows that none of the 12 secondary-search frequencies satisfies all of these criteria. The four-sigma amplitude scale of the search is about 0.016~mmag in the total system light. We therefore detect no stable, isolated companion clock in this frequency range. This does not exclude lower-amplitude oscillations, blended modes, or modes with phases that are not stable enough for pulsation timing.

\section{Stellar and Binary Modelling}\label{sec:model}

\subsection{MESA and GYRE grid}

\begin{figure*}
 \centering
 \includegraphics[width=0.95\textwidth]{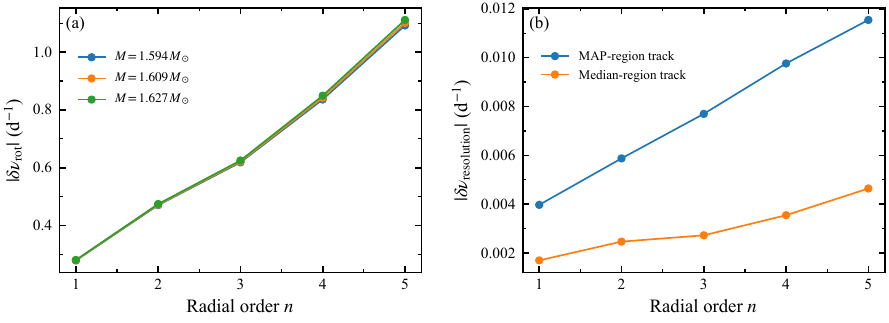}
 \caption{Model-frequency sensitivity tests.  (a) Absolute radial-frequency differences between matched rotating and nonrotating structures at three masses near the solution.  The rms shift across the three comparisons defines the systematic for each mode.  (b) Absolute changes from recomputing two tracks at higher temporal and spatial resolution.  The numerical changes remain below 0.012~d$^{-1}$.}
 \label{fig:systematics}
\end{figure*}

We calculated evolutionary tracks with Modules for Experiments in Stellar Astrophysics (MESA) r24.08.1 \citep{Paxton2011,Paxton2013,Paxton2015,Paxton2018,Paxton2019,Jermyn2023} and adiabatic radial-mode frequencies with GYRE 7.2.1 \citep{Townsend2013}. The main grid contains 172 tracks over $1.585\leq M/M_\odot\leq1.98$, $0.014\leq Z\leq0.022$, $1.70\leq\alpha_{\rm MLT}\leq2.10$, and $0.010\leq f_{\rm ov}\leq0.030$. We adopt the helium prescription used by \citet{Murphy2023},
\begin{equation}
Y=0.28+1.4\,(Z-0.0142),
\end{equation}
with the solar mixture of \citet{Asplund2009}, the \texttt{pp\_and\_cno\_extras.net} reaction network, and exponential core overshooting with $f_0=0.002$. We tested the dependence on the adopted helium relation by shifting the initial helium abundance by $\Delta Y=-0.01$ and $+0.01$ at three representative locations in the posterior region. The resulting changes in the radial frequencies and stellar properties were interpolated over the grid and propagated through the stellar, binary-SED, and Gaia-covariance calculation. The tracks run from the pre-main sequence through the young main sequence. We later added 38 independent tracks over $1.50\leq M/M_\odot\leq1.585$ to test the lower boundary. For the GYRE and SED calculations, we selected 30 main-sequence structures from each track within 1.25~d$^{-1}$ of the observed $\dnu$. The final fit therefore uses 6300 models from 210 tracks over a combined mass range of $1.50$--$1.98\,M_\odot$. No spectroscopic metallicity prior is applied. The metallicity is sampled within the model grid and constrained by the radial frequencies and the two-star SED.

The stellar likelihood uses the five candidate radial frequencies and the unresolved binary SED. We use the observed $\dnu$ to identify the ridge and select evolutionary models for the grid. Since it comes from the same pulsation spectrum, it is not included as a sixth independent seismic constraint, and no further $\dnu$ restriction is applied during posterior sampling. A complete third-order polynomial interpolates the GYRE frequencies over mass, composition, mixing length, core overshooting, and central hydrogen mass fraction. Track-grouped five-fold cross-validation gives mode-specific radial-frequency RMSEs no larger than about 0.12~d$^{-1}$, which are added in quadrature to the baseline mode uncertainties. We tested the normalization of the baseline model-comparison scales by repeating the complete inference after multiplying all five values by 0.75 and by 1.50. The interpolation errors, rotational terms, stellar grid, and all nonseismic constraints were unchanged.

We did not apply an empirical solar-like surface correction \citep[e.g.,][]{BallGizon2014} because the fitted modes are low-order $\delta$~Scuti modes and such corrections are not calibrated in this regime. Instead, we tested the sensitivity to the outer-layer treatment at three representative locations in the high-probability region. We recomputed the tracks using the baseline Eddington relation and the Krishna--Swamy and solar-Hopf $T(\tau)$ relations \citep{KrishnaSwamy1966,Paxton2013}. We also recalculated the radial-mode frequencies of the same reference structures using the JCD, VACUUM, and ISOTHERMAL GYRE outer boundary conditions \citep{Townsend2013}. The MESA atmosphere choices changed an individual fitted frequency by at most 0.067~d$^{-1}$, with a maximum rms change of 0.050~d$^{-1}$ across the five modes. Changing the GYRE outer boundary gave a maximum frequency change of 0.0034~d$^{-1}$. Both effects are smaller than the adopted baseline model-comparison uncertainties.

The reported $v\sin i=94\pm10$~km~s$^{-1}$ motivates including rotation in the frequency analysis. This value was measured from unresolved light, and absorption lines from the secondary may affect the inferred broadening. We therefore use it only to set the scale of the rotation tests; it is not included as a likelihood constraint. We selected three representative models near the inferred solution. At each location, we calculated a nonrotating model and rotating counterparts that reach surface velocities near 64, 96, and 127~km~s$^{-1}$. The middle sequence is closest to the measured $v\sin i$. Figure~\ref{fig:systematics} shows the frequency shifts for this sequence at the three representative locations. For the baseline likelihood, we adopt the rms shift across the three models near 96~km~s$^{-1}$ as the rotational uncertainty for each radial order. These terms range from 0.28 to 1.10~d$^{-1}$ for $n=1$--5. We also repeated the full inference for each rotation sequence, applying the changes in all five frequencies and in the stellar structure together. The baseline rotational uncertainty terms were omitted from these tests to avoid counting rotation twice. These calculations test the sensitivity to rotation but do not constitute a complete rotating grid.

For the radial-order comparison, we assigned the five frequencies in turn to $n=1$--5, $n=2$--6, and $n=3$--7. For each assignment, we evaluated all 6300 exact grid models using the same mode uncertainties. We then repeated the comparison after applying the order-dependent frequency shifts derived from rotating models with surface velocities of 92--98~km~s$^{-1}$.

\subsection{Physical binary SED}

The Gaia and Two Micron All Sky Survey (2MASS) photometry contains the unresolved light of both components. A single-star SED would therefore assign part of the companion light to the pulsator. We instead model the summed flux of two coeval stars. For each primary MESA model, the secondary mass is set to $M_2=qM_1$, where $q=K_1/K_2$ is determined by the Gaia SB2 orbit. At the same age and metallicity, the secondary properties are interpolated from the MESA Isochrones and Stellar Tracks (MIST) models \citep{Choi2016,Dotter2016}. Synthetic magnitudes for both components are calculated using the same MIST bolometric-correction grid and combined in flux. The MIST tracks are used only to describe the companion and the combined SED; the primary structure and pulsation frequencies are calculated with MESA and GYRE. The uncertainty in $q$ is propagated with the full Gaia orbital covariance in the final inference described in Section~\ref{sec:likelihood}.

The photometric likelihood uses Gaia $G$, $G_{\rm BP}$, and $G_{\rm RP}$ and 2MASS $JHK_s$, while parallax and extinction are sampled simultaneously. The Gaia astrometric fit has a renormalized unit weight error (RUWE) of 7.0, indicating a poor single-source astrometric fit and consistent with the known unresolved binary. We therefore inflate the formal Gaia parallax uncertainty by a factor of five and combine it with the Hipparcos new-reduction parallax \citep{vanLeeuwen2007}. The factor of five is a pragmatic choice rather than a calibrated correction. We tested Gaia uncertainty inflation factors of 3 and 10 and also repeated the inference using the Gaia and Hipparcos parallaxes separately. The Gaia SB2 orbit and its covariance were retained in all of these parallax tests. Using the formal Gaia uncertainty gives the same preferred grid mass, whereas omitting the Hipparcos constraint allows a broader high-mass tail. The SED constraints depend on the MESA--MIST combination, bolometric corrections, and adopted astrometric-error treatment; the quoted intervals do not include cross-grid or atmosphere-model systematics.

\begin{figure}
 \centering
 \includegraphics[width=0.43\textwidth]{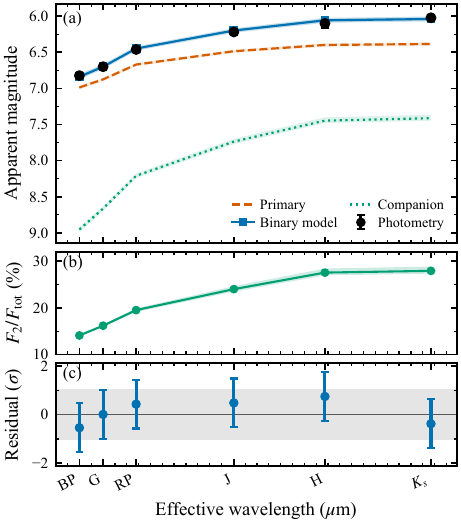}
 \caption{Two-component SED of \target.  (a) Unresolved Gaia and 2MASS photometry compared with the total binary model; dashed and dotted curves show the primary and companion.  Blue bands give the 68\% posterior interval.  (b) Companion fraction of the model passband flux.  (c) Photometric residuals normalized by the adopted uncertainties; the gray region marks $\pm1\sigma$.}
 \label{fig:sed}
\end{figure}

Figure~\ref{fig:sed} shows the resulting decomposition. The companion contributes about 14\% of the $G_{\rm BP}$ light, 16\% in $G$, 20\% in $G_{\rm RP}$, and 28\% in $K_s$. Its increasing fractional contribution toward the infrared reflects its lower effective temperature.

\subsection{Likelihood and Gaia covariance}\label{sec:likelihood}

The posterior for the stellar and orbital parameters has the form
\begin{equation}
\begin{aligned}
&p(\boldsymbol{\theta}_\star,\boldsymbol{\theta}_{\rm orb},
\varpi,A_V\mid
\boldsymbol{\nu},\boldsymbol{m},D_\varpi,\boldsymbol{D}_{\rm G})\\
&\quad\propto
p(\boldsymbol{\nu}\mid\boldsymbol{\theta}_\star)\,
p(\boldsymbol{m}\mid
\boldsymbol{\theta}_\star,
q(\boldsymbol{\theta}_{\rm orb}),
\varpi,A_V)\\
&\qquad\times
p(D_\varpi\mid\varpi)\,
p(\boldsymbol{D}_{\rm G}\mid\boldsymbol{\theta}_{\rm orb})\,
p(\boldsymbol{\theta}_\star)\,
p(\varpi)\,
p(A_V),
\end{aligned}
\end{equation}
where $\boldsymbol{\nu}$ contains the five radial frequencies, $\boldsymbol{m}$ contains the six unresolved apparent magnitudes, $D_\varpi$ contains the Gaia and Hipparcos parallax measurements, and $\boldsymbol{D}_{\rm G}$ denotes the Gaia NSS orbit. The sampled parallax and extinction are denoted by $\varpi$ and $A_V$. The stellar parameters have uniform priors within the model grid, the parallax is restricted to the adopted bounds, and the extinction has a non-negative half-normal prior. The Gaia orbital likelihood is represented by the published orbital solution and correlation vector.

We evaluate this posterior in two steps. A differential-evolution Markov chain Monte Carlo calculation first samples the stellar and SED parameters, including $\varpi$ and $A_V$, at the nominal Gaia mass ratio, using 32 walkers for 24,000 steps and discarding the first 8000 steps. The Gaia projected-mass constraint is excluded at this stage. All Gelman--Rubin statistics are below 1.002, and the minimum approximate effective sample size is 8459.

We then draw orbit vectors from the full seven-parameter Gaia covariance. Each draw supplies both $q=K_1/K_2$ for the binary SED and the projected component masses. We interpolate the SED likelihood over $q$ and reweight the stellar--orbit pairs by the change relative to the nominal-$q$ SED. Pairs for which the orbit-implied projected primary mass exceeds the stellar primary mass are rejected. This importance-sampling step has an effective sample size of about 40,800; an independent replicate changes the reported medians by less than 0.07 of their quoted $1\sigma$ uncertainties. The Gaia orbit therefore enters the calculation once, and the scalar SB2 projected mass is not included as an additional constraint.

Gaia provides the correlations among $P$, $e$, $K_1$, $K_2$, the systemic velocity, argument of periastron, and epoch of periastron passage through the NSS correlation vector. We reconstruct the covariance matrix and draw SB2 orbits. For each draw,
\begin{align}
M_1\sin^3 i &=
\frac{P(1-e^2)^{3/2}}{2\pi G}
(K_1+K_2)^2K_2,\\
M_2\sin^3 i &=
\frac{P(1-e^2)^{3/2}}{2\pi G}
(K_1+K_2)^2K_1.
\end{align}
For every retained stellar--orbit pair, we derive $i$ and $M_2=qM_1$. The nominal orbital elements give $M_1\sin^3 i\simeq1.589\,M_\odot$, but the Gaia covariance and the physical boundary produce an asymmetric allowed projected-mass distribution. We therefore propagate the full distribution rather than treating the nominal value of $1.589\,M_\odot$ as a fixed lower bound. 

Before applying the Gaia projected-mass constraint, the seismic and binary-SED fit at the nominal Gaia mass ratio gives $M_1=1.581^{+0.027}_{-0.025}\,M_\odot$. The final posterior gives $M_1=1.594^{+0.026}_{-0.026}\,M_\odot$. Including the Gaia orbital covariance shifts the median upward by about $0.014\,M_\odot$, while the width of the 68\% credible interval remains nearly unchanged. The seismic and SED fit therefore already provides a primary-mass constraint of comparable precision.

\begin{figure*}
\centering
\includegraphics[width=0.97\textwidth]{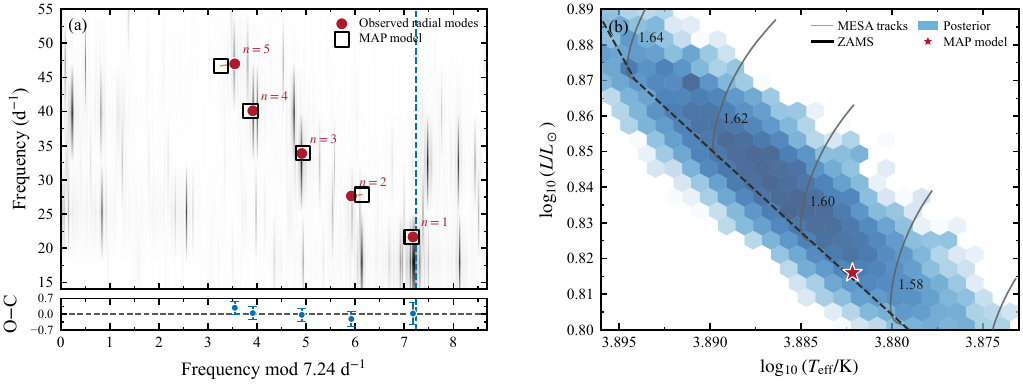}
\caption{Seismic and evolutionary solution. (a) Amplitude spectrum folded modulo $\Delta\nu=7.24$~d$^{-1}$, showing the observed radial modes (red circles) and the exact-grid maximum-a-posteriori (MAP) model frequencies (open squares). The lower strip shows the observed-minus-calculated (O$-$C) residuals, with vertical bars indicating the adopted mode uncertainties. (b) Zoomed HR diagram. Blue hexagons show the stellar posterior after propagation of the Gaia covariance, the red star marks the exact-grid MAP model, gray curves show local MESA tracks from the ZAMS to 400~Myr, and the dashed black curve marks the ZAMS. Track labels give the mass in $M_\odot$.}
\label{fig:evolution}
\end{figure*}

\section{Results}\label{sec:results}

Figure~\ref{fig:evolution} compares the five observed radial-mode candidates with the model predictions. All residuals are within 0.3~d$^{-1}$ and lie within the adopted model uncertainties. They are much larger than the formal Fourier errors but smaller than the frequency shifts in the rotating comparison models. Table~\ref{tab:mode_assignment} compares the three radial-order assignments quantitatively. In the nonrotating grid, the $n=2$--6 and $n=3$--7 assignments are worse than $n=1$--5 by $\Delta\chi^2=43.83$ and 323.12, respectively. After applying the order-dependent frequency shifts derived from the rotating models, the corresponding differences are 37.96 and 196.74. Rotation changes the size of these differences but not their ranking. The tested models therefore favor $n=1$--5.

\begin{deluxetable}{ccc}
\tablecaption{$\chi^2$ for Radial-order Assignments
\label{tab:mode_assignment}}
\setlength{\tabcolsep}{8pt}
\tablehead{
\colhead{Assignment} &
\colhead{Nonrotating} &
\colhead{With rotational shifts}
}
\startdata
$n=1$--5 & 0.93   & 2.50\\
$n=2$--6 & 44.76  & 40.46\\
$n=3$--7 & 324.05 & 199.24\\
\enddata
\end{deluxetable}

The exact-grid MAP model shown in Figure~\ref{fig:evolution} gives consecutive period ratios of 0.7764, 0.8217, 0.8462, and 0.8580. Their absolute differences from the observed ratios are 0.0018--0.0071, supporting the adopted radial sequence. The pulsating primary has $M_1=1.594^{+0.026}_{-0.026}\,M_\odot$, $R_1=1.482^{+0.014}_{-0.012}\,R_\odot$, $\teff,1=7699^{+78}_{-76}$~K, and $L_1=6.95^{+0.32}_{-0.30}\,L_\odot$. Its surface gravity is $\log g=4.299^{+0.006}_{-0.007}$. The solution lies close to the zero-age main sequence (ZAMS), with an age of $171^{+112}_{-99}$~Myr and a central hydrogen mass fraction near 0.67. The age is less precise than the mass and radius because young main-sequence tracks evolve slowly through this part of the Hertzsprung--Russell (HR) diagram. The inferred metallicity is $Z=0.0164^{+0.0013}_{-0.0015}$. This is a model-dependent result from the pulsation and SED fit, not an independent spectroscopic abundance measurement.

The HR panel of Figure~\ref{fig:evolution} is constrained by the radial frequencies and the two-star SED; we do not assign an unresolved catalogue temperature to the primary. The posterior distribution is shown against local MESA tracks from 1.50 to $1.70\,M_\odot$ in steps of $0.02\,M_\odot$, with the remaining inputs fixed near the posterior center. The tracks extend from the ZAMS to 400~Myr. The axes show only the region around the solution, so only the portions of the tracks crossing this window are visible. These tracks illustrate the mass ordering and direction of evolution but are not the grid used in the fit.

When the rotational changes are applied together, the sequence with a surface velocity near 96~km~s$^{-1}$ changes the median primary mass by $-0.0002\,M_\odot$, the radius by $-0.0019\,R_\odot$, and the age by $-40$~Myr. Across all three rotation sequences, the largest absolute changes are $0.0052\,M_\odot$, $0.0022\,R_\odot$, and 80~Myr, respectively. The primary mass and radius are therefore stable under the tested rotational changes, whereas the age is more sensitive to the treatment of rotation.

Changing the initial helium abundance by $\Delta Y=-0.01$ shifts the median primary mass, radius, and age by $+0.0124\,M_\odot$, $+0.0034\,R_\odot$, and $+5$~Myr, respectively. For $\Delta Y=+0.01$, the corresponding shifts are $-0.0121\,M_\odot$, $-0.0029\,R_\odot$, and $-12$~Myr. The largest mass and radius shifts are 0.48 and 0.27 times their statistical uncertainties, while the age shift is small compared with its wider posterior interval. Multiplying all five baseline model-frequency scales by 0.75 or 1.50 changes the median primary mass by less than $0.00011\,M_\odot$, the radius by less than $0.0006\,R_\odot$, and the age by less than 1~Myr. These shifts are below 0.05 times their corresponding statistical uncertainties. The stellar solution is therefore insensitive to the tested normalization of the model-frequency scales.

\begin{figure*}
 \centering
 \includegraphics[width=0.97\textwidth]{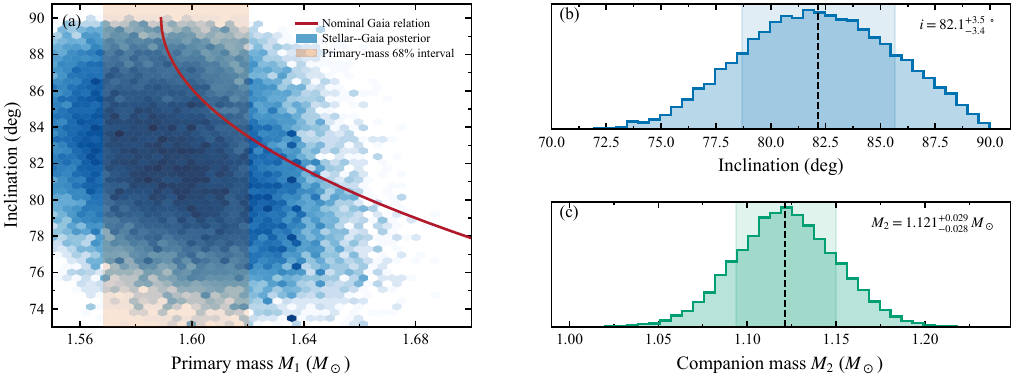}
 \caption{Stellar and dynamical solution. (a) Dark blue hexagons show the final stellar--Gaia posterior in the $M_1$--$i$ plane. Pale blue hexagons show the coverage of all 6300 MESA models after mapping through physically allowed Gaia covariance draws; this layer represents grid coverage rather than probability. The orange band marks the 16th--84th percentile interval of $M_1$, and the red curve shows the nominal relation $M_1\sin^3 i=1.589\,M_\odot$. Panels (b) and (c) show the inclination and companion-mass distributions; dashed lines mark the medians and shaded regions the 16th--84th percentiles.}
 \label{fig:bridge}
\end{figure*}

After propagating the Gaia covariance, we obtain $i=82.1^{+3.5}_{-3.4}$~deg and $M_2=1.121^{+0.029}_{-0.028}\,M_\odot$.  Figure~\ref{fig:bridge} shows how these quantities depend on the primary mass.  For comparison, the nominal value $M_1\sin^3i=1.589\,M_\odot$ combined with the median primary mass would give $i\simeq87.4^\circ$.  The final distribution differs from this single-value estimate because the correlated Gaia orbit, the physical boundary, and the $q$-dependent binary SED are propagated together.  The red nominal relation in Figure~\ref{fig:bridge} is therefore not expected to pass through the center of the full distribution.  The coeval SED gives $\teff,2=5947^{+89}_{-98}$~K, $R_2=1.030^{+0.041}_{-0.035}\,R_\odot$, and $L_2=1.19^{+0.14}_{-0.11}\,L_\odot$.  These values describe a late-F or early-G dwarf below the main $\delta$~Scuti instability region. 

With the Hipparcos parallax retained, changing the Gaia uncertainty inflation factor from 3 to 10 changes the median primary mass by no more than $0.0023\,M_\odot$. The Hipparcos-only parallax solution also remains close to the adopted result. In contrast, the Gaia-only parallax solution gives $M_1=1.615^{+0.034}_{-0.034}\,M_\odot$, a radius higher by $0.0077\,R_\odot$, and an age lower by 34~Myr. Its 95th-percentile mass increases from $1.636$ to $1.672\,M_\odot$. The adopted central solution is therefore insensitive to the exact Gaia inflation factor, but the narrower high-mass tail depends mainly on the Hipparcos parallax.

\begin{deluxetable*}{llccc}
\tabletypesize{\small}
\setlength{\tabcolsep}{9pt}
\tablecaption{Sensitivity of the Primary Parameters
\label{tab:sensitivity}}
\tablehead{
\colhead{Test} &
\colhead{Cases} &
\colhead{Maximum $|\Delta M_1|$} &
\colhead{Maximum $|\Delta R_1|$} &
\colhead{Maximum $|\Delta{\rm Age}|$}\\
\colhead{} &
\colhead{} &
\colhead{($M_\odot$)} &
\colhead{($R_\odot$)} &
\colhead{(Myr)}
}
\startdata
Atmosphere & Krishna--Swamy, solar--Hopf
& 0.0001 & 0.0006 & 4.2\\
Rotation & 64--127~km~s$^{-1}$
& 0.0052 & 0.0022 & 80\\
Helium abundance & $\Delta Y=\pm0.01$
& 0.0124 & 0.0034 & 12\\
Frequency scales & $0.75$--$1.50$ times baseline
& 0.0001 & 0.0006 & 1.0\\
Gaia uncertainty & $3$--$10$ times, with Hipparcos
& 0.0023 & 0.0008 & 6.8\\
Gaia-only parallax & Gaia uncertainty times 5
& 0.0208 & 0.0077 & 34\\
Hipparcos-only parallax & Gaia parallax omitted
& 0.0002 & 0.0005 & 2.4\\
\enddata
\tablecomments{Entries are the largest absolute changes relative to the adopted calculation over the cases listed in the second column. Gaia-only and Hipparcos-only refer to the parallax likelihood; the Gaia SB2 orbit is retained in both cases.}
\end{deluxetable*}

Table~\ref{tab:sensitivity} summarizes the main sensitivity tests. The largest change in the primary mass occurs when the Hipparcos parallax is omitted, while rotation has the largest effect on the inferred age. These tests examine specific alternatives rather than sample a probability distribution over the uncertain model physics. Their parameter shifts are therefore reported separately and are not added in quadrature to the posterior intervals.

\begin{deluxetable}{lcccc}
\tabletypesize{\small}
\setlength{\tabcolsep}{6pt}
\tablecaption{Adopted Stellar and Binary Parameters\label{tab:parameters}}
\tablehead{
\colhead{Parameter} &
\colhead{Median} &
\colhead{$-1\sigma$} &
\colhead{$+1\sigma$} &
\colhead{Basis}
}
\startdata
$M_1$ ($M_\odot$) & 1.594 & 0.026 & 0.026 & Full fit\\
$R_1$ ($R_\odot$) & 1.482 & 0.012 & 0.014 & Full fit\\
$\teff,1$ (K) & 7699 & 76 & 78 & Full fit\\
$L_1$ ($L_\odot$) & 6.95 & 0.30 & 0.32 & Full fit\\
$\log g_1$ (cgs) & 4.299 & 0.007 & 0.006 & Seis.+SED\\
Age (Myr) & 171 & 99 & 112 & Full fit\\
$Z$ & 0.0164 & 0.0015 & 0.0013 & Full fit\\
$M_{\rm cc}$ ($M_\odot$) & 0.173 & 0.009 & 0.008 & MESA\\
$i$ (deg) & 82.1 & 3.4 & 3.5 & Full fit\\
$M_2$ ($M_\odot$) & 1.121 & 0.028 & 0.029 & Full fit\\
$\teff,2$ (K) & 5947 & 98 & 89 & Coeval SED\\
$R_2$ ($R_\odot$) & 1.030 & 0.035 & 0.041 & Coeval SED\\
$L_2$ ($L_\odot$) & 1.19 & 0.11 & 0.14 & Coeval SED
\enddata
\tablecomments{Uncertainties are posterior 16th--84th percentile intervals. Full-fit parameters include the propagation of the full Gaia orbital covariance and are conditional on the $n=1$--5 mode identification and adopted stellar physics. The secondary properties are derived from the coeval MIST model at the same age and metallicity as the primary, with the Gaia mass-ratio uncertainty propagated through the final inference.}
\end{deluxetable}

The selected models have a convective-core mass near $0.173\,M_\odot$.  Five radial modes constrain the global structure more strongly than the detailed chemical stratification, so we do not treat the internal abundance profiles as independent measurements.  The core mass in Table~\ref{tab:parameters} is therefore a model-dependent structural quantity rather than a direct seismic observable.

\section{Discussion}\label{sec:discussion}

Previous work identified the regular pulsation ridge, measured its spacing, proposed the fundamental radial mode, and obtained a single-star mass near $1.65\,M_\odot$ \citep{Bedding2020,Murphy2023,Panda2024}. Our mass of about $1.59\,M_\odot$ agrees with that estimate, but now refers specifically to the Gaia primary. The radial modes and two-star SED constrain this star, the timing delays identify it as the pulsator, and the SB2 orbit then gives the inclination and companion mass. The result is a physical description of both stars rather than a single-star fit to their combined light.

The primary is a young main-sequence star close to the ZAMS. Its central hydrogen mass fraction is about 0.67, so little hydrogen has been consumed in the core. The five low-order radial modes match the models to better than 1\% and mainly constrain the mean density and global structure. They do not constrain the detailed chemical profile or core overshooting with the same precision. This is why the mass and radius are better determined than the age.

The two components also occupy different pulsation regimes. The primary is near 7700~K, where $\delta$~Scuti pulsation is expected, whereas the companion is near 5950~K and lies below the main $\delta$~Scuti temperature range. This temperature difference naturally explains why the regular pulsation pattern is associated with the primary. The absence of a stable, anti-phased clock from the companion is consistent with its lower temperature, although it does not rule out weaker or less stable variability.

The binary SED is important because the cooler companion contributes more light at longer wavelengths. Its contribution rises from about 14\% in $G_{\rm BP}$ to about 28\% in $K_s$. A single-star SED would assign this light to the pulsator and could bias the inferred primary properties. Constant companion light dilutes the observed pulsation amplitudes but does not change a stable pulsation frequency or phase. The radial-mode pattern and timing delays are therefore preserved, while a weak companion mode becomes harder to detect.

The stellar mass also clarifies the Gaia measurements. The value near $1.49\,M_\odot$ in the Gaia \texttt{binary\_masses} table is not a dynamical component mass and is smaller than the minimum primary mass implied by the nominal SB2 orbit. The latter gives $M_1\sin^3 i\simeq1.589\,M_\odot$. Combined with our inferred primary mass of $1.594\,M_\odot$, the nominal orbit alone would imply $i\simeq87.4^\circ$. After propagating the full Gaia covariance, the physical boundary, and the $q$-dependent binary SED, the posterior instead gives $i=82.1^{+3.5}_{-3.4}$~deg. The separations at the two conjunctions are approximately 55.7 and $64.7\,R_\odot$, while the sum of the stellar radii is about $2.51\,R_\odot$. Eclipses would require inclinations above approximately $87.4^\circ$ and $87.8^\circ$, respectively. The single-value estimate lies close to the lower limit, whereas the posterior median is below both limits, consistent with the absence of eclipses in the TESS photometry.

Pulsation timing and the frequency fit serve different purposes. The five radial modes constrain the stellar structure, whereas the two isolated timing modes identify the pulsating component. The gaps between TESS campaigns leave the timing-only orbit aliased, so we do not use it to replace the Gaia orbit. Its value here is the agreement in timing sign and amplitude with the motion predicted for the Gaia primary.

Related binary studies have used eclipses, red-giant oscillations, or solar-like modes to set the mass scale \citep{GarciaHernandez2017,Murphy2021,Beck2024}. HD~70510 instead uses a sequence of low-order radial modes in the $\delta$~Scuti primary of a non-eclipsing SB2. It provides a component-specific test of a young pulsation model and transfers that mass scale to a cooler companion without a detected timing clock.

Rotation has the largest effect on the inferred age among the model-physics variations tested here. The rotating-model tests change the median age by up to 80~Myr, while leaving the primary mass and radius nearly unchanged. If the published $v\sin i$ applies to the primary and its spin is aligned with the orbit, the equatorial velocity is about 95~km~s$^{-1}$, corresponding to a rotation period near 0.8~d, much shorter than the 42-d orbital period. However, no peak in the expected 1.10--1.45~d$^{-1}$ range recurs across the four TESS observing epochs, so the rotation period is not independently measured. The primary parameters remain conditional on the $n=1$--5 identification, the adopted helium enrichment relation, and the adopted one-dimensional models, while the companion properties depend on the coeval MIST models and bolometric corrections. Multi-epoch high-resolution spectroscopy could test the solution by independently measuring both velocity amplitudes and providing component-specific temperatures, metallicities, and an optical light ratio.

The same approach can be applied beyond this system. TESS provides regular mode sequences and stable clocks, while Gaia NSS supplies SB2 orbits for many non-eclipsing binaries. The most useful targets will have a clear radial-mode sequence and sufficient photometry to constrain the component fluxes. Such systems can provide component masses for testing $\delta$~Scuti models without requiring eclipses.

\section{Conclusions}\label{sec:conclusions}

We identify five high-S/N frequencies in \target\ as the radial fundamental mode and four consecutive overtones. The radial modes and two-star SED, combined with the Gaia SB2 orbit, give $M_1=1.594^{+0.026}_{-0.026}\,M_\odot$, $R_1=1.482^{+0.014}_{-0.012}\,R_\odot$, and an age of $171^{+112}_{-99}$~Myr. The primary is a young $\delta$~Scuti star close to the ZAMS.

Two stable pulsation clocks follow the motion expected for the Gaia primary and identify it as the pulsating component. The companion is a cooler, roughly 6000-K dwarf that supplies 14--28\% of the unresolved optical-to-near-infrared flux. We find no stable, anti-phased clock from the companion, consistent with its position below the main $\delta$~Scuti temperature range, although weaker or less stable variability remains possible.

Propagating the full Gaia SB2 orbital covariance with the stellar constraints gives $i=82.1^{+3.5}_{-3.4}$~deg and $M_2=1.121^{+0.029}_{-0.028}\,M_\odot$. The quoted intervals are statistical and remain conditional on the $n=1$--5 identification, the adopted one-dimensional stellar physics, and the binary-SED construction. The sensitivity tests leave the primary mass and radius stable, while the inferred age is more sensitive to rotation.

HD~70510 shows that regular radial modes can set the mass scale in a non-eclipsing SB2. Pulsation timing assigns the pulsating component to the Gaia primary, and the orbit then transfers the stellar mass scale to the companion. Applying the same method to regular TESS pulsators with Gaia NSS orbits can provide component masses for tests of young $\delta$~Scuti models.

\begin{acknowledgments}
Chenglong Lv acknowledges support from the National Natural Science Foundation of China (NSFC; No. 12503042) and the Natural Science Foundation of Xinjiang Uygur Autonomous Region (No. 2025D01B176). Ali Esamdin is supported by the Tian-shan Talent Training Program (No. 2023TSYCLJ0053). This work is also supported by the Xinjiang Uygur Autonomous Region Introduced Project "Tianchi Talent".

This paper includes data collected with the \textit{TESS} mission, obtained from the Mikulski Archive for Space Telescopes (MAST) at the Space Telescope Science Institute (STScI). Funding for the \textit{TESS} mission is provided by the NASA Explorer Program. STScI is operated by the Association of Universities for Research in Astronomy, Inc., under NASA contract NAS~5-26555. The specific TESS SPOC 2-min observations analyzed in this work can be accessed from MAST via \dataset[doi:10.17909/ebb2-8r93]{https://doi.org/10.17909/ebb2-8r93}. This work also presents results from the European Space Agency (ESA) space mission Gaia. Gaia data are being processed by the Gaia Data Processing and Analysis Consortium (DPAC). Funding for the DPAC is provided by national institutions, in particular the institutions participating in the Gaia MultiLateral Agreement (MLA). The Gaia mission website is \url{https://www.cosmos.esa.int/gaia}. The Gaia archive website is \url{https://archives.esac.esa.int/gaia}. The MESA modeling files used in this work are available on Zenodo: \dataset[doi:10.5281/zenodo.23691991] {https://doi.org/10.5281/zenodo.23691991}.
\end{acknowledgments}

\software{MESA \citep{Paxton2011,Paxton2013,Paxton2015,Paxton2018,Paxton2019,Jermyn2023}, GYRE \citep{Townsend2013}, FELIX \citep{Charpinet2010}, Astropy \citep{Astropy2013,Astropy2018}, NumPy \citep{Harris2020}, SciPy \citep{Virtanen2020}, matplotlib \citep{Hunter2007}, pandas \citep{McKinney2010}}

\bibliographystyle{aasjournalv7}
\bibliography{references}

@ARTICLE{Bedding2020,
  author       = {{Bedding}, Timothy R. and {Murphy}, Simon J. and {Hey}, Daniel R. and others},
  title        = {Very regular high-frequency pulsation modes in young intermediate-mass stars},
  journal      = {Nature},
  year         = {2020},
  volume       = {581},
  pages        = {147--151},
  doi          = {10.1038/s41586-020-2226-8}
}

@ARTICLE{Lv2024,
  author       = {{Lv}, C. and {Esamdin}, A. and {Hasanzadeh}, A. and {Ghazinejad}, M. and {Pascual-Granado}, J. and {Mirouh}, G. M. and {Karimov}, R.},
  title        = {Statistical analysis of asteroseismic indices and stellar parameters of TESS-observed delta Scuti stars},
  journal      = {\aap},
  year         = {2024},
  volume       = {686},
  pages        = {A174},
  doi          = {10.1051/0004-6361/202346985}
}

@ARTICLE{Lv2022,
  author = {{Lv}, Chenglong and {Esamdin}, Ali and
            {Pascual-Granado}, J. and {Yang}, Taozhi and
            {Shen}, Dongxiang},
  title = "{Frequency Analysis of KIC 1573174:
            Shedding Light on the Nature of HADS Stars}",
  journal = {\apj},
  year = 2022,
  month = jun,
  volume = {932},
  number = {1},
  pages = {42},
  doi = {10.3847/1538-4357/ac69d9}
}

@ARTICLE{Murphy2023,
  author       = {{Murphy}, Simon J. and {Bedding}, Timothy R. and {Gautam}, Anuj and {Joyce}, Meridith},
  title        = {A grid of 200,000 models of young delta Scuti stars using MESA and GYRE},
  journal      = {\mnras},
  year         = {2023},
  volume       = {526},
  pages        = {3779--3795},
  doi          = {10.1093/mnras/stad2849}
}

@ARTICLE{Panda2024,
  author       = {{Panda}, Subrata Kumar and {Dhanpal}, Siddharth and {Murphy}, Simon J. and {Hanasoge}, Shravan and {Bedding}, Timothy R.},
  title        = {Asteroseismology Applied to Constrain Structure Parameters of Delta Scuti Stars},
  journal      = {\apj},
  year         = {2024},
  volume       = {960},
  number       = {2},
  pages        = {94},
  doi          = {10.3847/1538-4357/ad0a97}
}

@ARTICLE{GarciaHernandez2017,
  author       = {{Garc\'ia Hern\'andez}, A. and {Su\'arez}, J. C. and {Moya}, A. and others},
  title        = {Precise surface gravities of delta Scuti stars from asteroseismology},
  journal      = {\mnras},
  year         = {2017},
  volume       = {471},
  number       = {1},
  pages        = {L140--L144},
  doi          = {10.1093/mnrasl/slx117}
}

@ARTICLE{Murphy2014,
  author       = {{Murphy}, Simon J. and {Bedding}, Timothy R. and {Shibahashi}, Hiromoto and {Kurtz}, Donald W. and {Kjeldsen}, Hans},
  title        = {Finding binaries among Kepler pulsating stars from phase modulation of their pulsations},
  journal      = {\mnras},
  year         = {2014},
  volume       = {441},
  number       = {3},
  pages        = {2515--2527},
  doi          = {10.1093/mnras/stu765}
}

@ARTICLE{Murphy2018,
  author       = {{Murphy}, Simon J. and {Moe}, Maxwell and {Kurtz}, Donald W. and {Bedding}, Timothy R. and {Shibahashi}, Hiromoto and {Boffin}, Henri M. J.},
  title        = {Finding binaries from phase modulation of pulsating stars with Kepler. V. Orbital parameters, with eccentricity and mass-ratio distributions of 341 new binaries},
  journal      = {\mnras},
  year         = {2018},
  volume       = {474},
  number       = {4},
  pages        = {4322--4346},
  doi          = {10.1093/mnras/stx3049}
}

@ARTICLE{Murphy2021,
  author       = {{Murphy}, Simon J. and {Li}, Tanda and {Sekaran}, Sanjay and others},
  title        = {A binary with a delta Scuti star and an oscillating red giant: orbit and asteroseismology of KIC 9773821},
  journal      = {\mnras},
  year         = {2021},
  volume       = {505},
  number       = {2},
  pages        = {2336--2348},
  doi          = {10.1093/mnras/stab1436}
}

@ARTICLE{GaiaMultiplicity,
  author       = {{Gaia Collaboration} and {Arenou}, Fr\'ed\'eric and others},
  title        = {Gaia Data Release 3: Stellar multiplicity, a teaser for the hidden treasure},
  journal      = {\aap},
  year         = {2023},
  volume       = {674},
  pages        = {A34},
  doi          = {10.1051/0004-6361/202243782}
}

@ARTICLE{Beck2024,
  author       = {{Beck}, P. G. and {Grossmann}, D. H. and {Steinwender}, L. and others},
  title        = {Constraining stellar and orbital co-evolution through ensemble seismology of solar-like oscillators in binary systems: A census of oscillating red giants and dwarf stars in Gaia DR3 binaries},
  journal      = {\aap},
  year         = {2024},
  volume       = {682},
  pages        = {A7},
  doi          = {10.1051/0004-6361/202346810}
}

@ARTICLE{Ricker2015,
  author       = {{Ricker}, George R. and {Winn}, Joshua N. and {Vanderspek}, Roland and others},
  title        = {Transiting Exoplanet Survey Satellite (TESS)},
  journal      = {Journal of Astronomical Telescopes, Instruments, and Systems},
  year         = {2015},
  volume       = {1},
  number       = {1},
  pages        = {014003},
  doi          = {10.1117/1.JATIS.1.1.014003}
}

@ARTICLE{Paxton2011,
  author       = {{Paxton}, Bill and {Bildsten}, Lars and {Dotter}, Aaron and {Herwig}, Falk and {Lesaffre}, Pierre and {Timmes}, Frank},
  title        = {Modules for Experiments in Stellar Astrophysics (MESA)},
  journal      = {\apjs},
  year         = {2011},
  volume       = {192},
  number       = {1},
  pages        = {3},
  doi          = {10.1088/0067-0049/192/1/3}
}

@ARTICLE{Paxton2013,
  author       = {{Paxton}, Bill and {Cantiello}, Matteo and {Arras}, Phil and others},
  title        = {Modules for Experiments in Stellar Astrophysics (MESA): Planets, Oscillations, Rotation, and Massive Stars},
  journal      = {\apjs},
  year         = {2013},
  volume       = {208},
  number       = {1},
  pages        = {4},
  doi          = {10.1088/0067-0049/208/1/4}
}

@ARTICLE{Paxton2015,
  author       = {{Paxton}, Bill and {Marchant}, Pablo and {Schwab}, Josiah and others},
  title        = {Modules for Experiments in Stellar Astrophysics (MESA): Binaries, Pulsations, and Explosions},
  journal      = {\apjs},
  year         = {2015},
  volume       = {220},
  number       = {1},
  pages        = {15},
  doi          = {10.1088/0067-0049/220/1/15}
}

@ARTICLE{Paxton2018,
  author       = {{Paxton}, Bill and {Schwab}, Josiah and {Bauer}, Evan B. and others},
  title        = {Modules for Experiments in Stellar Astrophysics (MESA): Convective Boundaries, Element Diffusion, and Massive Star Explosions},
  journal      = {\apjs},
  year         = {2018},
  volume       = {234},
  number       = {1},
  pages        = {34},
  doi          = {10.3847/1538-4365/aaa5a8}
}

@ARTICLE{Paxton2019,
  author       = {{Paxton}, Bill and {Smolec}, R. and {Schwab}, J. and others},
  title        = {Modules for Experiments in Stellar Astrophysics (MESA): Pulsating Variable Stars, Rotation, Convective Boundaries, and Energy Conservation},
  journal      = {\apjs},
  year         = {2019},
  volume       = {243},
  number       = {1},
  pages        = {10},
  doi          = {10.3847/1538-4365/ab2241}
}

@ARTICLE{Jermyn2023,
  author       = {{Jermyn}, Adam S. and {Bauer}, Evan B. and {Schwab}, Josiah and others},
  title        = {Modules for Experiments in Stellar Astrophysics: Time-dependent Convection, Energy Conservation, Automatic Differentiation, and Infrastructure},
  journal      = {\apjs},
  year         = {2023},
  volume       = {265},
  number       = {1},
  pages        = {15},
  doi          = {10.3847/1538-4365/acae8d}
}

@ARTICLE{Asplund2009,
  author       = {{Asplund}, Martin and {Grevesse}, Nicolas and {Sauval}, A. Jacques and {Scott}, Pat},
  title        = {The Chemical Composition of the Sun},
  journal      = {\araa},
  year         = {2009},
  volume       = {47},
  pages        = {481--522},
  doi          = {10.1146/annurev.astro.46.060407.145222}
}

@ARTICLE{Choi2016,
  author       = {{Choi}, Jieun and {Dotter}, Aaron and {Conroy}, Charlie and {Cantiello}, Matteo and {Paxton}, Bill and {Johnson}, Benjamin D.},
  title        = {Mesa Isochrones and Stellar Tracks (MIST). I. Solar-scaled Models},
  journal      = {\apj},
  year         = {2016},
  volume       = {823},
  number       = {2},
  pages        = {102},
  doi          = {10.3847/0004-637X/823/2/102}
}

@ARTICLE{Dotter2016,
  author       = {{Dotter}, Aaron},
  title        = {MESA Isochrones and Stellar Tracks (MIST) 0: Methods for the Construction of Stellar Isochrones},
  journal      = {\apjs},
  year         = {2016},
  volume       = {222},
  number       = {1},
  pages        = {8},
  doi          = {10.3847/0067-0049/222/1/8}
}

@ARTICLE{vanLeeuwen2007,
  author       = {{van Leeuwen}, Floor},
  title        = {Validation of the new Hipparcos reduction},
  journal      = {\aap},
  year         = {2007},
  volume       = {474},
  pages        = {653--664},
  doi          = {10.1051/0004-6361:20078357}
}

@ARTICLE{Townsend2013,
  author       = {{Townsend}, R. H. D. and {Teitler}, S. A.},
  title        = {GYRE: an open-source stellar oscillation code based on a new Magnus Multiple Shooting scheme},
  journal      = {\mnras},
  year         = {2013},
  volume       = {435},
  pages        = {3406--3418},
  doi          = {10.1093/mnras/stt1533}
}

@ARTICLE{Astropy2013,
  author       = {{Astropy Collaboration} and {Robitaille}, Thomas P. and {Tollerud}, Erik J. and others},
  title        = {Astropy: A Community Python Package for Astronomy},
  journal      = {\aap},
  year         = {2013},
  volume       = {558},
  pages        = {A33},
  doi          = {10.1051/0004-6361/201322068}
}

@ARTICLE{Astropy2018,
  author       = {{Astropy Collaboration} and {Price-Whelan}, Adrian M. and {Sip\H{o}cz}, Brigitta M. and others},
  title        = {The Astropy Project: Building an Inclusive, Open-science Project and Status of the v2.0 Core Package},
  journal      = {\aj},
  year         = {2018},
  volume       = {156},
  number       = {3},
  pages        = {123},
  doi          = {10.3847/1538-3881/aabc4f}
}

@ARTICLE{Harris2020,
  author       = {{Harris}, Charles R. and {Millman}, K. Jarrod and {van der Walt}, St\'efan J. and others},
  title        = {Array programming with NumPy},
  journal      = {Nature},
  year         = {2020},
  volume       = {585},
  pages        = {357--362},
  doi          = {10.1038/s41586-020-2649-2}
}

@ARTICLE{Virtanen2020,
  author       = {{Virtanen}, Pauli and {Gommers}, Ralf and {Oliphant}, Travis E. and others},
  title        = {SciPy 1.0: Fundamental Algorithms for Scientific Computing in Python},
  journal      = {Nature Methods},
  year         = {2020},
  volume       = {17},
  pages        = {261--272},
  doi          = {10.1038/s41592-019-0686-2}
}

@ARTICLE{Hunter2007,
  author       = {{Hunter}, John D.},
  title        = {Matplotlib: A 2D Graphics Environment},
  journal      = {Computing in Science \& Engineering},
  year         = {2007},
  volume       = {9},
  number       = {3},
  pages        = {90--95},
  doi          = {10.1109/MCSE.2007.55}
}

@INPROCEEDINGS{McKinney2010,
  author       = {{McKinney}, Wes},
  title        = {Data Structures for Statistical Computing in Python},
  booktitle    = {Proceedings of the 9th Python in Science Conference},
  year         = {2010},
  pages        = {56--61},
  doi          = {10.25080/Majora-92bf1922-00a},
  editor       = {{van der Walt}, St\'efan and {Millman}, Jarrod}
}

@ARTICLE{Lv2023,
  author       = {{Lv}, Chenglong and {Esamdin}, Ali and {Hasanzadeh}, A. and {Ma}, Shuguo and {Yang}, TaoZhi and {Karimov}, Rivkat},
  title        = {{TESS Observations of Seven Newly Identified High-amplitude {\ensuremath{\delta}} Scuti Stars}},
  journal      = {\apj},
  year         = {2023},
  volume       = {959},
  number       = {1},
  pages        = {33},
  doi          = {10.3847/1538-4357/acf999}
}

@ARTICLE{Lv2026a,
  author       = {{Lv}, Chenglong and {Esamdin}, Ali and {Pascual-Granado}, J. and {Feng}, Guojie and {Wang}, Haozhi and {Karimov}, Rivkat and {Juraev}, Bolkin and {Ehgamberdiev}, Shuhrat A.},
  title        = {{Discovery and Characterization of Three New High-amplitude {\ensuremath{\delta}} Scuti Stars from TESS Observations}},
  journal      = {\apj},
  year         = {2026},
  volume       = {1002},
  number       = {1},
  pages        = {101},
  doi          = {10.3847/1538-4357/ae5dc5}
}

@ARTICLE{BallGizon2014,
  author = {{Ball}, W. H. and {Gizon}, L.},
  title = {A new correction of stellar oscillation frequencies for near-surface effects},
  journal = {A\&A},
  year = {2014},
  volume = {568},
  pages = {A123},
  doi = {10.1051/0004-6361/201424325}
}

@ARTICLE{KrishnaSwamy1966,
  author = {{Krishna Swamy}, K. S.},
  title = {Profiles of Strong Lines in K-Dwarfs},
  journal = {ApJ},
  year = {1966},
  volume = {145},
  pages = {174--194},
  doi = {10.1086/148752}
}

@ARTICLE{Charpinet2010,
  author       = {{Charpinet}, S. and {Green}, E.~M. and {Baglin}, A. and {Van Grootel}, V. and {Fontaine}, G. and {Vauclair}, G. and {Chaintreuil}, S. and {Weiss}, W.~W. and {Michel}, E. and {Auvergne}, M. and {Catala}, C. and {Samadi}, R. and {Baudin}, F.},
  title        = {{CoRoT opens a new era in hot B subdwarf asteroseismology. Detection of multiple g-mode oscillations in KPD 0629-0016}},
  journal      = {\aap},
  year         = {2010},
  volume       = {516},
  pages        = {L6},
  doi          = {10.1051/0004-6361/201014789}
}

@ARTICLE{Lv2026b,
       author = {{Lv}, Chenglong and {Pascual-Granado}, J. and {Esamdin}, Ali and {Gardini}, Angela and {Wang}, Haozhi and {Yao}, Jianwei and {Ma}, Shuguo},
        title = "{Asteroseismic Detection of a Massive Unseen Companion to the {\ensuremath{\delta}} Scuti Star TIC 160582982}",
      journal = {\apj},
         year = 2026,
        month = oct,
       volume = {1010},
       number = {2},
          eid = {199},
        pages = {199},
          doi = {10.3847/1538-4357/aea560},
       adsurl = {https://ui.adsabs.harvard.edu/abs/2026ApJ..1010..199L}
}

@ARTICLE{Lv2026c,
       author = {{Lv}, Chenglong and {Esamdin}, Ali and {Ma}, Shuguo and {Yaqup}, Shahidin and {Wang}, Haozhi and {Jmahali}, Ayzada and {Yao}, Jianwei and {Kamili}, Kedieryemu},
        title = "{TESS Pulsation Timing Reveals the JWST Calibration Star HD 158485 as a Double-pulsator Binary}",
      journal = {\apjl},
         year = 2026,
        month = nov,
       volume = {1011},
       number = {1},
          eid = {L20},
        pages = {L20},
          doi = {10.3847/2041-8213/aeb301},
archivePrefix = {arXiv},
       eprint = {2610.14371},
 primaryClass = {astro-ph.SR},
       adsurl = {https://ui.adsabs.harvard.edu/abs/2026ApJ..1011L..20L}
}

\end{document}